# Topology of photonic time-crystals

**Eran Lustig[†], Yonatan Sharabi[†] and Mordechai Segev**

*Solid State Institute and Physics Department*

*Technion – Israel Institute of Technology, Haifa 32000, Israel.*

[†]These authors contributed equally to this work

Abstract:

We introduce topological phases in Photonic Time-Crystals. Photonic Time-Crystals are materials in which the refractive index varies periodically and abruptly in time. When the refractive index changes abruptly, the light propagating in the material experiences time-refraction and time-reflection, analogous to refraction and reflection in photonic crystals. Interference between time-refracted and time-reflected waves gives rise to Floquet-Bloch states and dispersion bands, which are gapped in the momentum. We show that photonic time-crystals can be in a topologically non-trivial phase, and calculate the topological invariant associated with the momentum bands of the Photonic Time-Crystal. The topological invariants are related to the phase between the forward and backward-propagating waves of the time-crystal, and to localized edge states in time.

The past decade has seen the rapid development of topological phases of matter in many areas of physics. These phases classify the structure of the electronic bands of materials[1], and are robust to the presence of disorder or defects. The study of topological phases led to the discovery of new and exciting materials such as topological insulators[2]. However, topological phases are not only a class of electronic condensed matter systems. Rather, they are a universal wave phenomenon that manifests itself in optics[3], cold-atoms[4,5], acoustics[6,7], exciton-polaritons [8,9] and more. Specifically, in optics, nontrivial topological phases were demonstrated to produce topological edge states [10], robust unidirectional light propagation [11–13], high efficient lasing [14,15], Thouless pumping [16] and other exotic phenomena [17,18].

An interesting type of crystals that was not related thus far to topological phases is the Photonic Time-Crystal (PTC)[19–23]. PTCs are the temporal analogues of photonic crystals. While photonic crystals are designed with a refractive index $n(r)$ that varies periodically in space, one can think of a PTC where the refractive index changes periodically in time: $n(t)$.

Due to the unique duality of space and time in Maxwell's equations, certain photonic-crystals and PTCs are analogous to one another [21]. Consequently, a sudden temporal change in the permittivity $\epsilon$ causes time reflections similar to a sudden change of $\epsilon$ in space, causing spatial reflections [24–26]. Time reflections that occur in a periodic manner lead to interference between forward propagating waves and time-reversed waves, giving rise to Floquet-Bloch states and dispersion bands, which are gapped in the momentum $k$, rather than in frequency[21,22]. In fact, band-gaps in momentum can occur for any periodic modulation of the refractive index $n(t)$[22]. Generally, temporal modulation of the refractive index is extremely useful for many purposes such as realizing optical isolators without magnetic fields, stopping light and creating synthetic gauge fields for light [27–30]. Importantly, for observing these phenomena even relatively small changes in the refractive index suffice, as was demonstrated for silicon photonics [31,32].

Small changes in refractive index cause small band-gaps in momentum, and therefore in most of these cases the momentum gaps are not substantial. However, to observe the pronounced features of a PTC, the modulation frequency must be high enough and the amplitude large enough such that the momentum gaps are substantial, and have the same order of magnitude as the momentum of the light propagating in the PTC (otherwise, for small momentum gaps the observed effects are inconsequential). Thus far, PTCs were only demonstrated at radio frequencies in electronic transmission lines [33]. However, the transition to optical frequencies is close, as PTCs are now attracting growing attention due to recent advances in fabricating dynamic optical systems and metamaterials and are expected to be observed in the near future [34–36]. In fact, systems with permittivity that varies on short time-scales were demonstrated in many optical systems. Nonlinear effects such as the Kerr nonlinearity in epsilon near zero materials [36], and plasma generation [37] can temporally change the permittivity and refractive index of a material at very fast time scales and at large amplitudes. Specifically, the promising results in [36] show that time-modulated materials with large modulation amplitudes in the refractive index are now within fabrication capabilities of modern technology, and that epsilon near zero materials are good candidates to create a PTC in the very near future.

Here, we introduce topological phases in PTCs. We describe PTCs displaying a non-trivial topology, which forms by periodically modulating any homogeneous dielectric medium in specific fashion that can be as simple as periodic step-like modulation of the refractive index. We show that PTCs have a topology akin to that of topological insulators. We prove analytically and demonstrate in FDTD simulations, that the topological invariant of the dispersion bands in momentum is related to the relative phase between the forward and backward propagating waves generated by the PTC. The topology also gives rise to "temporal topological edge-states", which are the temporal analogue of topological edge-states. A temporal edge-state is manifested by a localized peak in the amplitude around the time of the temporal edge-state. This peak is formed by the interface between two different PTCs, with exponentially decaying tails in both temporal directions.

The system we analyze is a spatially homogeneous material with permittivity $\epsilon(t)$, which is modulated in time, such that $\epsilon(t)$ changes periodically, with period $T$, in a step-like manner. This results in a binary PTC with two time segments. In the first time-segment $\epsilon(t) = \epsilon_1$ for a duration of $t_1$ seconds, followed by a second time-segment in which $\epsilon(t) = \epsilon_2$ for $t_2 = T - t_1$ seconds (Fig.1a). For simplicity, the field is polarized in the $x$ direction, and propagates in the $z$ direction. This model was pioneered in [22], which showed that this system yields a PTC. Here, we address the topological features of this system, and its implications in photonics and in topological physics. With every modulation of $\epsilon(t)$, a time reflection occurs, causing waves to partially reflect to their time reversed pair, while preserving the momentum k due to the homogeneity of space. The time reversed partner of a wave is a wave with the same momentum but with opposite spatial frequency. This is analogous to a wave conserving its energy, and scattering backwards in space in a photonic crystal.

To derive the field propagation in the PTC, we start by describing a linearly polarized wave with momentum $k$ incident on the PTC. In each time-segment the frequency is proportional to the momentum: $\omega_\alpha = k n_\alpha c$, where $\alpha$ is the segment number and is equal to 1 or 2, $n_\alpha = \sqrt{\epsilon_\alpha}$, $k$ is the vacuum wavenumber of the wave, and $c$ is the speed of light in vacuum.

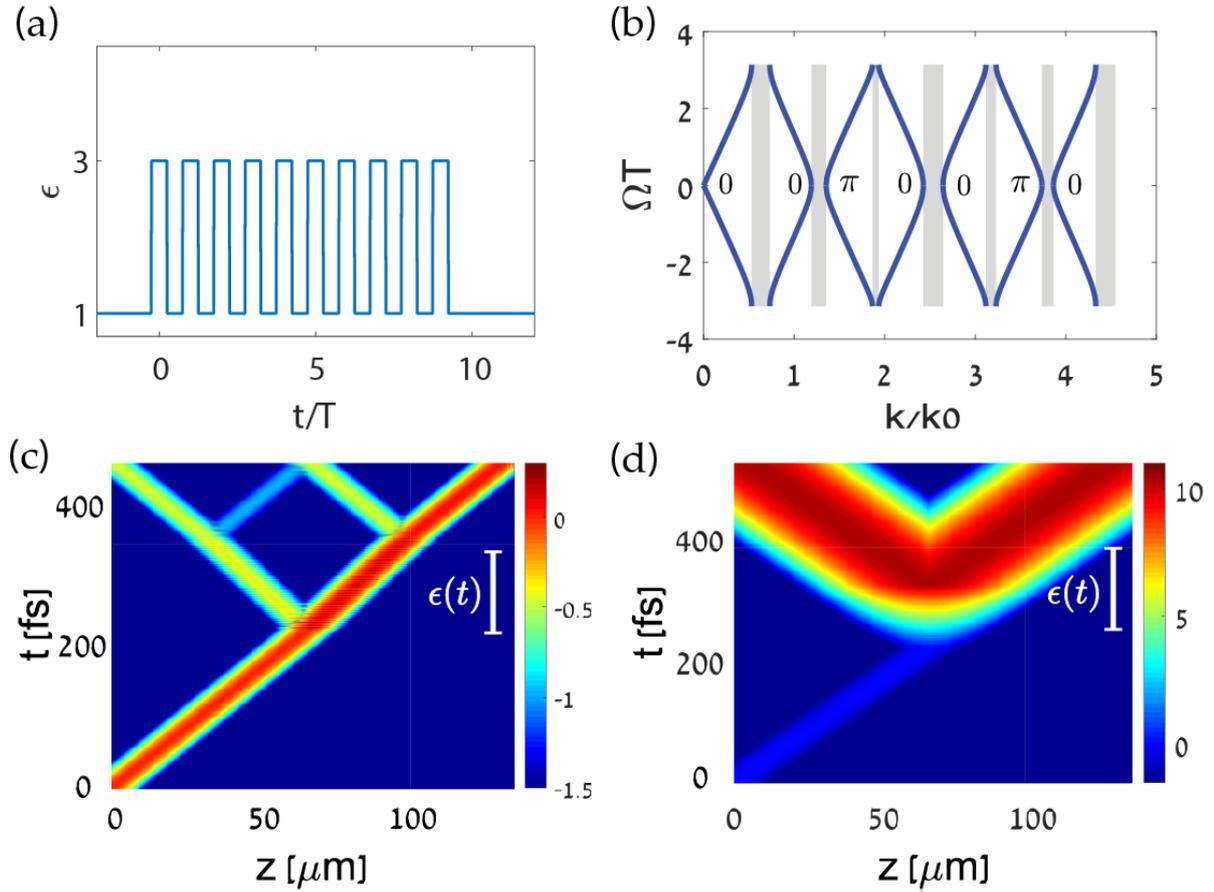

Fig. 1. (a) A binary temporal photonic crystal (PTC), for $\epsilon_1 = 3$ and $\epsilon_2 = 1$ (b) Dispersion bands of the PTC (blue lines) separated by gaps (gray regions) in momentum. The values (0 and $\pi$) labeling each band are the Zak phases associated with the band. (c-d) FDTD simulations, showing the field amplitude (in Log scale) of a pulse propagating under the influence of a PTC. The PTC takes place during the time-period marked in white. In (c) the pulse resides in a momentum band, while in (d) it resides in a momentum band gap. Consequently, the pulse in (c) undergoes two splitting events: when it enters the PTC and when it leaves the PTC, and eventually four pulses emerge from the PTC. On the other hand, in (d) the pulse undergoes a single splitting into two pulses.

Thus, the displacement field takes the form:

$$D_x^{(\alpha,n)} = \left(a_n^{(\alpha)} e^{i\omega_\alpha\left(t+\frac{t_1}{2}-nT\right)} + b_n^{(\alpha)} e^{-i\omega_\alpha\left(t+\frac{t_1}{2}-nT\right)}\right) e^{-ikz}, \tag{1}$$

where $D_x^{(\alpha,n)}$ is the electric displacement field in time period $n$ and time segment $\alpha$, and, $a_n^{(\alpha)}$, $b_n^{(\alpha)}$ are complex amplitudes (complex numbers) representing the field amplitudes times the permittivity in each time segment. Notice that we choose $t = 0$ to be in the middle of time segment 1. This will turn out to be convenient since it perseveres time reversal symmetry. In the same fashion, the magnetic field $B_y$ has a similar form (related through the waves' impedance). The electric displacement field $D_x$ and the magnetic field $B_y$ are continuous between different time segments. Imposing this continuity results in the matrix equation:

$$\begin{pmatrix} a_n^{(\alpha)} \\ b_n^{(\alpha)} \end{pmatrix} = \begin{pmatrix} W^{(\alpha)} & -Y^{(\alpha)} \\ -Z^{(\alpha)} & X^{(\alpha)} \end{pmatrix}^n \begin{pmatrix} a_0^{(\alpha)} \\ b_0^{(\alpha)} \end{pmatrix}. \tag{2}$$

The explicit form of $W^{(\alpha)}, X^{(\alpha)}, Y^{(\alpha)}, Z^{(\alpha)}$ is given in[38]. We also note that $W^{(1)} + X^{(1)} = W^{(2)} + X^{(2)} \equiv W + X$.

Next, we turn to find the band structure, and the eigenstates of the infinite PTC. According to the Floquet theorem for each momentum component k, the displacement field assumes the form $D_x = D_\Omega(t)e^{-i\Omega t}e^{-ikz}$ where $D_\Omega(t) = D_\Omega(t+T)$. Imposing the Floquet form on Eq.(2) yields the Floquet dispersion relation:

$$\Omega(k) = \frac{1}{T}\cos^{-1}(W+X), \tag{3}$$

and the Floquet states:

$$\begin{pmatrix} a_n^{(1)} \\ b_n^{(1)} \end{pmatrix} = e^{-i\Omega nt} \begin{pmatrix} Y^{(1)} \\ e^{i\Omega T} - X^{(1)} \end{pmatrix}, \tag{4}$$

where $a_n^{(2)}$ and $b_n^{(2)}$ are obtained by imposing the continuity conditions on $a_n^{(1)}$ and $b_n^{(1)}$. We plot the Floquet frequency $\Omega$ as a function of the momentum $k$ in Fig.1b, which constitutes the band structure of the PTC. The values of $k$ for which $\Omega$ is real are the bands (blue lines in Fig.1b), and the band-gaps are the regions in which $\Omega$ is complex (grey regions in Fig.1b).

In many 1D systems, the topological invariant is given by the Zak phase[39]. By analogy, one can formulate the Zak phase of each PTC band as:

$$\theta_m^{Zak} = \int_{-\frac{\pi}{T}}^{\frac{\pi}{T}} d\Omega \left[ i\int_0^T dt \epsilon(t) u_{m,\Omega}^*(t) \partial_\Omega u_{m,\Omega}(t) \right], \tag{5}$$

where $m$ is the band index and $\theta_m^{Zak}$ is the Zak phase. We find that the Zak phase takes the values of zero or $\pi$ for each band. The proof that these are the only possible values for the Zak phase, follows similar lines to the proof for the Zak phase of the spatial photonic crystal in [40], and is given in detail in [38]. As a concrete example, we calculate the Zak phases of the bands of a PTC with $\epsilon_1 = 3, \epsilon_2 = 1, t_1 = t_2 = 0.5T, T = 2[fs]$, and present them in Fig.1b near the appropriate bands.

The main question at this point is what features are dictated by the Zak phase for the light propagating in the PTC. These features are interesting since they are topological, and therefore should be robust to defects and disorder. The study of these properties is done analytically and verified in simulation by numerically solving Maxwell's equations with the finite difference time domain method (FDTD).

To describe the effect of the topology on the PTC, we first need to describe in more details the characteristic behavior of light inside and outside a band-gap in momentum. To do so, we simulate the propagation of two different pulses inside the PTC. Both pulses have a full width at half maximum of $5[fs]$. While the first pulse has a center wavelength of $1.4[um]$ that falls within a band, the other has a center wavelength of $0.93\ [um]$ that falls within the bandgap. At $t = 0$ the pulse starts propagting in free space ($\epsilon = 1$), and at time $t = 220[fs]$ a PTC with the same parameters as before starts. The system is linear, therefore we can decompose the pulse propagation to the propagation of monochromatic plane waves, and study it according to Eq.(1-4).

According to Eq.(1-4), after the PTC starts each plane-wave component of the pulse couples to two Floquet-modes, one propagating in the positive z direction (forward propagating), and the other in the negative z direction (time reversed). Figure 1c shows the displacement field amplitude of a pulse for which the momentum components reside in the bands of the PTC. Since Ω(k) is real in the bands, the intensity of the pulse remains constant on average during propagation. When the PTC ends at $t_{end} = 340[fs]$ after $n = 60$ periods, the two Floquet-modes split again, each coupling back to a forward and time reversed plane wave. As a result, four different pulses exist after the time crystal ends.

On the other hand, if the momentum components of the pulse fall within the band gap, $\Omega(k)$ takes a complex value that results in an exponential increase of energy as time progresses (Fig 1.d). When the time crystal begins, the pulse becomes localized in space, with an exponentially increasing amplitude. Once the time crystal ends, the pulse couples to plane waves, but this time the result is two pulses instead of four for the case of propagation in the band. The two pulses are seen in Fig.1d after the PTC ends. The topological properties of the PTC dictates the phase between these two pulses as we explain next.

It is instructive to compare the effect of the PTC on a pulse to the effect of a spatial photonic crystal. Consider again a 1D spatial photonic crystal: a 1D system of equally spaced layers of alternating refractive indices. When light is incident on this 1D photonic crystal and has frequencies in the band-gap of the spatial photonic crystal, all the light is reflected (Fig.2a). In this case the incident wave and the reflected wave have the same amplitude but different phases. The Zak phase dictates the sign of the phase between the incident wave $E_i$ and the reflected wave $E_r$ [40]. In this case, the lattice behaves as a perfect mirror ($E_t = 0$) and the light does not enter the photonic-crystal if its frequency falls within a bandgap. Considering now the temporal equivalent, we find that in a PTC the situation is fundamentally different (Fig.2b). A wave cannot travel back in time, and therefore all the light enters the PTC, regardless of its momentum – even if it falls within the band gap. This calls for a new interpretation of the Zak phase for PTCs. It turns out that, in a PTC, the Zak phase dictates the sign of the phase between the transmitted wave and the reflected wave, when their momentum is inside the band-gap.

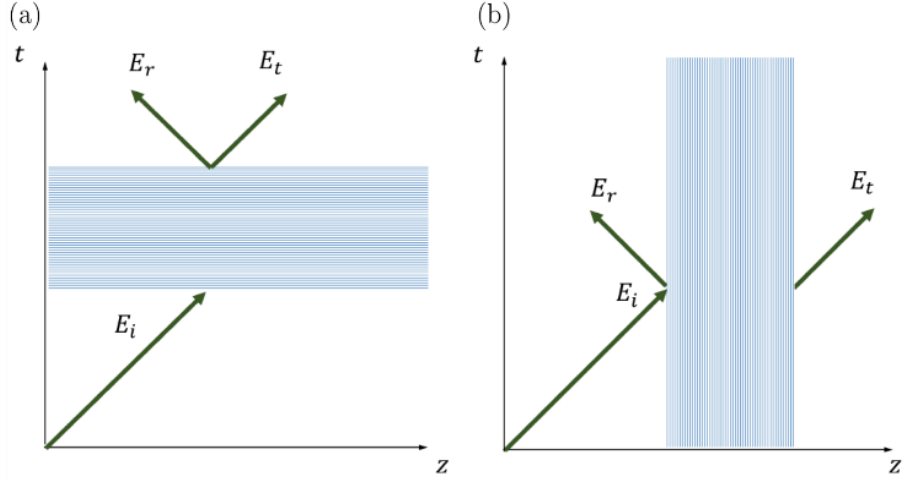

Fig. 2. Reflection and transmission schematics for a monochromatic plane wave incident upon a PTC - (a), compared to an ordinary 1D photonic crystal in (b). In the photonic crystal, if $E_i$ is in a band gap, $E_t$ is zero, whereas for a PTC the light always enters the time-crystals and passes through it, even if the wavenumber falls within a bandgap.

We now demonstrate how the Zak phase dictates the phase difference between the transmitted wave and the reflected wave in a PTC. Consider an incident plane wave $E_i e^{i\omega t}$ experiencing $n$ periods of a PTC (Fig.1a). The plane wave has wavenumber k inside a band gap. After the PTC ends, the pulse splits into a time-refracted wave: $tE_i e^{i\omega t}$, and a time reflected wave $rE_i e^{-i\omega t}$ as depicted in Fig.1d. It is evident that in this case $|t| = |r|$ [41], but the relative phase $t/r = e^{i\phi}$ is different in each band-gap and for each k. The sign of $\phi$ is determined by the Zak phase according to:

$$\text{sgn}(\phi_s) = \delta(-1)^{s+l} \exp\left(i \sum_{m=1}^{s-1} \theta_m^{Zak}\right), \quad (6)$$

where s is the gap number (lowest gap number is 1), $\delta = \text{sgn}(1 - \epsilon_a \mu_b / \epsilon_b \mu_a)$ and l is the number of band crossings below gap s. For the first six gaps in our example, $\text{sgn}(\phi_1) = \text{sgn}(\phi_2) = \text{sgn}(\phi_6) = 1$ and $\text{sgn}(\phi_3) = \text{sgn}(\phi_4) = \text{sgn}(\phi_5) = -1$. To derive Eq.(6), we compare the displacement field $D_x$ divided by the magnetic field $B_y$ at $t_{end}^+$ with the same quantity at time $t_{end}^-$ [38]. This gives the relation:

$$\frac{t+r}{t-r} = c\mu_0 n_\alpha \frac{Y^{(1)}e^{i\omega_\alpha t^*} + \left(e^{i\Omega T} - X^{(1)}\right)e^{-i\omega_\alpha t^*}}{Y^{(1)}e^{i\omega_\alpha t^*} - \left(e^{i\Omega T} - X^{(1)}\right)e^{-i\omega_\alpha t^*}} \tag{7}$$

where $t^* = t_{end} + t_1/2 - nT$ and $\alpha$ is 1 or 2 according to the last layer in the PTC. From Eq.(7) the phase between $t$ and r can be retrieved [38]. The relative phase between two pulses has significant influence on the observables of many systems. For example, coherent control pump-probe experiments, and electromagnetically induced transparency depend on this quantity. Furthermore, for short pulses the relative phase can have an even greater importance as the carrier-envelope phase holds information about the distribution and maximal value of the electric field.

To verify the phase signs obtained analytically from topological considerations (the Zak phase) for the first six gaps, we calculate (numerically) the relative phases between the Floquet modes in the FDTD simulation of Fig.1d. We find the phase of each frequency component by Fourier transforming the fields, and then calculating the difference between the frequency component of the time-refracted and the time-reversed Floquet modes. The phase differences are plotted in Fig.3(a-f) for the first six bands. The sign of ϕ calculated from our FDTD simulation exactly matches the analytic relation found in Eq.(6). This proves that indeed the relative phase sign between the time-refracted and the time-reversed fields in this PTC is determined uniquely by the topological properties of the system.

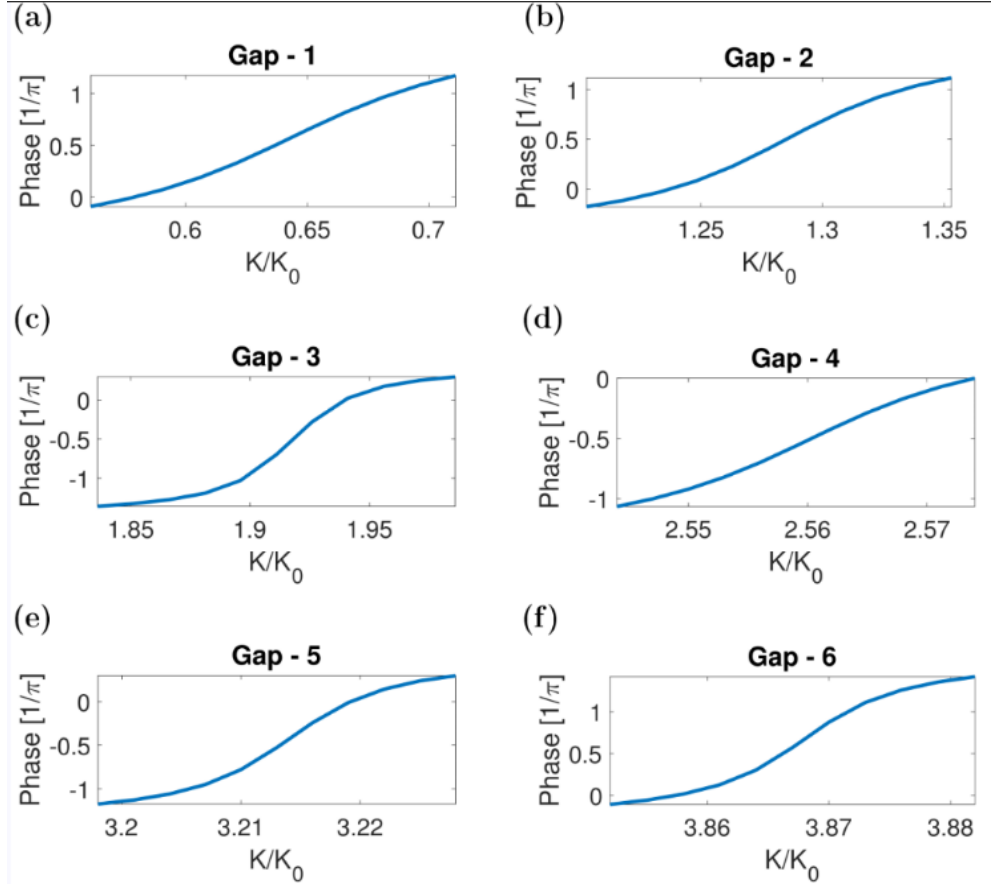

Fig. 3. (a-f) Phase difference $\phi$ between the forward and backward propagating Floquet modes of the first six gaps of the PTC, obtained from FDTD simulations. These results match the calculation based on the Topological invariant - the Zak phase, in Eq. (5-6).

Topological phases govern the propagation of light in a PTC in other ways besides the phase just studied. As known from topological physics, when placing two sub-systems with different topological phases next to each other, edge states appear at the interface between them. The edge states are eigen-states of the entire system, confined to the region between the two sub-systems. In this context, we ask the following question: what is an edge-state in time?

To answer this question, we study the dynamics of light in a system composed of two sequential PTCs, shown in Fig. 4a. The first PTC has $\epsilon_1 = 3, \epsilon_2 = 1$ and $t_1 = t_2 = 0.5T$, while the second PTC has $\epsilon_2 = 3, \epsilon_1 = 1$ and $t_1 = t_2 = 0.5T$ (Fig 4.a). Thus, the two PTCs have the exact same band gaps, and yet a different topology, manifested in the different Zak phases of their bands. Such two PTCs occur immediately one after the other, as demonstrated in Fig 4.a, where the interface is a specific time point $t_{edge} = 8T$. We find that a topological edge state forms at the temporal interface between two PTCs. Such an edge-state in time is essentially a temporal exponential increase in amplitude towards $t_{edge}$ followed by a temporal exponential decrease. The peak amplitude of this process is at $t_{edge}$. After the light amplitude decreases, it begins to increase once again. We plot the energy of the light in the PTC as a function of time near the temporal edge-state in Fig. 4b.

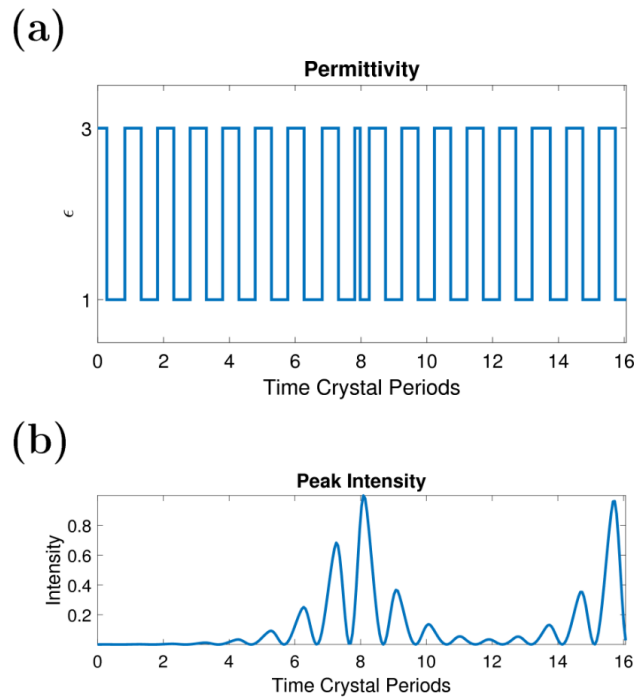

Fig. 4. Temporal topological edge states between two PTCs. (a) Two PTCs with different Zak phases cascaded at $t = 8T$. (b). Intensity of the wave propagating in the PTC plotted in (a). The light intensity increases exponentially in time up to the interface between PTCs at $t = 8T$. Immediately after the interface, the intensity starts to decrease exponentially, before eventually returning to its normal band-gap behavior of an exponential increase as the time progresses. The result is an intensity peak around the time of the interface between PTCs.

It is instructive to compare our results on temporal topological edge states between two PTCs and spatial topological edge-states between two crystals. In the spatial case, the topological edge states always exhibit exponential decay on both sides of the interface. Topological edge-states of time crystals on the other hand are between regions that each supports exponential increase of the intensity. Thus, the phenomenon observed in PTCs of a peak decaying in both temporal directions is completely counterintuitive. This decrease in intensity on either side of the topological temporal edge state between two PTCs means that the external modulation extracts energy from the system in a lossless manner, which can have many implications.

Last but not least, we propose a realization for the topological PTC with an epsilon-near-zero material based on Al-doped zinc oxide. Very recently, it was demonstrated that the refractive index of this material can be set to vary uniformly in time at optical frequencies [36]. The experiment created a reflected time-reversed wave with high efficiency, thus fulfilling the requirements for the temporal topological effects presented here. This system is an excellent candidate for exploring topological effects in PTCs. We also stress that regardless of its interesting properties, measuring the topological properties of a PTC is also a useful approach for studying time crystals in experiments. The reason is that topological phenomena are fundamentally robust to disorder, and therefore topological lattices produce exact measurements even when fabricated with low accuracy, as is typically the case in pioneering experiments. Furthermore, the topology here is applicable for the entire electromagnetic spectrum, thus it can also be demonstrated even at radio frequencies in electronic transmission lines [33]

Before closing, we would like to emphasize the similarities and differences between the topological PTC presented here and another class of topological systems arising from temporal modulation: Floquet topological insulators [42], which were demonstrated in photonics [12]. The topology we present here is fundamentally different from Floquet topological insulators. In Floquet topological insulators the driving field, which can be a temporal (or spatial) modulation, is auxiliary designed for opening a topological gap in the frequency dispersion $\omega(k)$ of a crystal, that is, in a system that is periodic in space. In contradistinction, for PTCs, the time is the crystalline dimension itself. Hence, the abrupt temporal variations in PTCs open a topological gap in the momentum $k$ and not in the frequency $\omega$. This is a property of systems governed by equations with a second derivative in time, such as the wave equation in electromagnetism. Consequently, the ideas proposed here are relevant to virtually any wave system in nature, where the properties of the medium can be varied in time.

In conclusion, we studied a topological photonic time-crystal, proved that it has distinct topological phases, and demonstrated how they affect the propagation of light inside the time-crystal. Understanding the topological phenomena associated with PTCs can lead to many interesting avenues of research. For example, many interesting topological phenomena appear only in dimensions higher than one. Here, even though there is only one time dimension, adding a PTC to spatial one-dimensional lattices (such as the SSH lattice) would lead to the formation of topological insulators in space-time. Finally, we note that PTCs also have interesting quantum properties[43], as they are related to photon pair production from the vacuum state and to squeezed light [44]. Undoubtedly, experimental avenues for realizing PTCs are now rapidly developing, and experiments with PTCs are expected in the near future. Unraveling the topological aspects of PTCs, along with their classical and quantum properties would greatly help in understanding the fundamental nature of time, light, and periodic phenomena.

# Topology of Photonic Time-Crystals: Supplementary Material

Here we give the mathematical details of the topology of the photonic time-crystal. In Fig.5 we illustrate the photonic time-crystal (PTC). It is a binary PTC with two repeating segments. In the first time-segment $\epsilon(t) = \epsilon_1$ for a duration of $t_1$ seconds, followed by a second time-segment in which $\epsilon(t) = \epsilon_2$ for $t_2 = T - t_1$ seconds. For simplicity, the field is polarized in the $x$ direction, and propagates in the $z$ direction.

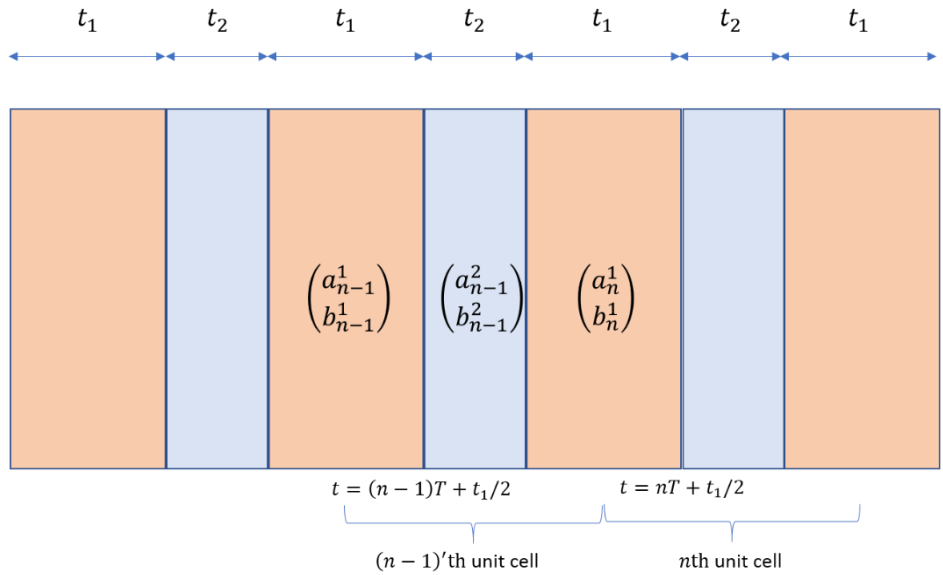

Fig. 5: The time lattice. Each color corresponds to a different refractive index and the horizontal direction is time.

We approach the problem with the transfer matrix formalism [45] .The continuity conditions on a temporal boundary at $t = t_0$ between two time segments is given by:

$$\boldsymbol{D}(x, t = t_0^+) = \boldsymbol{D}(x, t = t_0^-), \tag{8}$$

$$\boldsymbol{B}(x, t = t_0^+) = \boldsymbol{B}(x, t = t_0^-), \tag{9}$$

where $B$ is the magnetic field, and $D$ is the displacement field. We fix the time axis such that the system will be symmetric to time inversion: $t \to -t$. $t = 0$ is set to be at the middle of a segment $t_1$ (See Fig.5). The time inversion symmetry is required for defining the topological invariants. The momentum $k$ of a wave is conserved due to homogeneity of space, and during each time-segment the light has a dispersion relation:

$$\omega_\alpha = k n_\alpha c \tag{10}$$

where $\alpha = 1,2$ and $n_\alpha$ is the refractive index of layer $\alpha$. Under these constraints the electric displacement of a wave with momentum $k$ is given by:

$$D_x(t,z) = \left[ a_n^{(\alpha)} e^{-i\omega_\alpha \cdot \left(t + \frac{t_\alpha}{2} - nT\right)} + b_n^{(\alpha)} e^{i\omega_\alpha \cdot \left(t + \frac{t_\alpha}{2} - nT\right)} \right] e^{ik_z z}, \tag{11}$$

where $a_n^{(\alpha)}, b_n^{(\alpha)}$ are constant coefficients for each $n, \alpha$. The magnetic field can be obtained through Maxwell's equations in each segment (away from the boundaries):

$$\nabla \times \mathbf{B} = \frac{1}{\mu_0} \frac{\partial \mathbf{D}}{\partial t}, \tag{12}$$

$$ik_z B_y = i\frac{\omega_\alpha}{\mu_0}\left[ -a_n^{(\alpha)} e^{i\omega_\alpha \cdot \left(t + \frac{t_\alpha}{2} - nT\right)} + a_n^{(\alpha)} e^{-i\omega_\alpha \cdot \left(t + \frac{t_\alpha}{2} - nT\right)} \right], \tag{13}$$

where $\mu_0$ is the magnetic susceptibility. At this point, it will be convenient to use the notations:

$$a_n^{(1)} \equiv a_n, \qquad b_n^{(1)} \equiv b_n, \qquad a_n^{(2)} \equiv c_n, \qquad b_n^{(2)} \equiv d_n, \tag{14}$$

Substituting equations (12) and (13) into the continuity conditions yields the transfer matrices for $a_n, b_n, c_n, d_n$:

$$\begin{pmatrix} a_{n-1} \\ b_{n-1} \end{pmatrix} = \begin{pmatrix} A_1 & B_1 \\ C_1 & D_1 \end{pmatrix} \begin{pmatrix} a_n \\ b_n \end{pmatrix}, \tag{15}$$

$$\begin{pmatrix} c_{n-1} \\ d_{n-1} \end{pmatrix} = \begin{pmatrix} A_2 & B_2 \\ C_2 & D_2 \end{pmatrix} \begin{pmatrix} c_n \\ d_n \end{pmatrix}, \tag{16}$$

where $A_\alpha = e^{i\omega_\alpha t_\alpha}\left[\cos(\omega_{\bar\alpha} t_{\bar\alpha}) + \frac{1}{2}i\left(\frac{\omega_2}{\omega_1} + \frac{\omega_1}{\omega_2}\right)\cos\omega_{\bar\alpha} t_{\bar\alpha}\right]$,

$B_\alpha = \frac{i}{2} e^{-i\omega_\alpha(T(\alpha)+t_\alpha)}\left(\frac{\omega_{\bar\alpha}}{\omega_\alpha} - \frac{\omega_\alpha}{\omega_{\bar\alpha}}\right)\sin\omega_{\bar\alpha} t_{\bar\alpha}$, $C_\alpha = B_\alpha^*$, $D_\alpha = A_\alpha^*$, $T(\alpha) = \begin{cases} 0 & \alpha = 1 \\ T & \alpha = 2 \end{cases}$ and

$$\bar{\alpha} = \begin{cases} 2 & \alpha = 1 \\ 1 & \alpha = 2 \end{cases}$$

Equations (15) and (16) give the fields $D$ and $B$ for any initial conditions. However, the topological properties are defined for the infinite system. Since $\epsilon(t)$ is periodic in time, according to Floquet-Bloch theorem, the displacement field is of the form:

$$D(t) = D_\Omega(t) e^{-i\Omega t} e^{-ik_z z}, \tag{17}$$

where $D_\Omega(t)$ is periodic with a period $T$. Applying condition (17) on Eqs. (15) and (16) gives the Floquet dispersion:

$$\Omega(k_z) = \frac{1}{T} \cos^{-1}(D + A), \tag{18}$$

And the eigenmodes:

$$\begin{pmatrix} a_0^\alpha \\ b_0^\alpha \end{pmatrix} = \begin{pmatrix} B_\alpha \\ e^{i\Omega T} - A_\alpha \end{pmatrix}, \tag{19}$$

where we used the fact that: $D + A = D_\alpha + A_\alpha$. Thus, in the infinite lattice the displacement field is given by:

$$D_x(t,z) = \left[ a_0^{(\alpha)} e^{-i\omega_\alpha \cdot \left(t + \frac{t_\alpha}{2} - nT\right)} + b_0^{(\alpha)} e^{i\omega_\alpha \cdot \left(t + \frac{t_\alpha}{2} - nT\right)} \right] e^{-i\Omega nT} e^{ik_z z}. \tag{20}$$

The bands are where $\Omega$ takes real values and the gaps are where $\Omega$ takes imaginary values.

Next, we define the Zak phase for the displacement field $D$. The Zak phase is given by:

$$\theta_m^{Zak} = i \int_{-\frac{\pi}{T}}^{\frac{\pi}{T}} d\Omega \int_{temporal\, period} dt\, D_x(t,z) \partial_\Omega D_x(t,z), \tag{21}$$

where $m$ is the band number. Since the system has inversion symmetry ($t \to -t$), $\theta_m^{Zak}$ is quantized to be 0 or $\pi$ [39]

Next, we will show that the phase between the two emerging pulses of the PTC is calculated through the Zak phase. Consider an incoming plane wave $D_{in} = D_0 e^{-i\omega t}$. The reflected wave from the PTC is $D_R = RD_0 e^{i\omega t}$ and the transmitted wave is $D_T = TD_0 e^{-i\omega t}$. We take $\omega$ to be inside bandgap number $s$. A moment before the PTC ends, at time $t = nT^-$, the electric displacement field is described by the Floquet-Bloch state in Eq.(20). To find the expression for the phase between $D_T$ and $D_R$ use the continuity of $D$ and $B$:

$$\frac{D_x(nT^+)}{B(nT^+)} = \frac{D_x(nT^-)}{B(nT^-)}, \qquad (22)$$

Substituting the Floquet-Bloch modes on the left hand side of Eq.(22) and the reflection and transmission coefficients on the right hand side gives, after some algebraic manipulation:

$$\frac{1+\frac{R}{T}}{1-\frac{R}{T}} = \frac{\mu_0}{\omega_\alpha k_z} \frac{B_\alpha e^{-i\omega_\alpha nT} + (e^{i\Omega T} - A_\alpha) e^{i\omega_\alpha nT}}{B_\alpha e^{-i\omega_\alpha nT} - (e^{i\Omega T} - A_\alpha) e^{i\omega_\alpha nT}}. \qquad (23)$$

We define the phase $\phi_s$ to be the phase between the refracted wave and the transmitted wave remaining after the PTC ends. Since $R$ and $T$ are equal in amplitude we have:

$$\frac{R}{T} = e^{i\phi_s}. \qquad (24)$$

From Eq.(23) and (24) it is possible [40] to calculate the phase, and show that the phase sign of $\phi_s$ is related to the Zak phase according to:

$$\text{sgn}(\phi_s) = \delta(-1)^s (-1)^l \exp\left(i \sum_{m=1}^{s-1} \theta_m^{Zak}\right), \qquad (25)$$

where $s$ is the gap number (lowest gap number is 1), $\delta = \text{sgn}(1 - \epsilon_a\mu_b/\epsilon_b\mu_a)$ and l is the number of metallic band crossings below gap $s$.